
\input harvmac

%
%
%
%
\ifx\answ\bigans
\else
\output={
  \almostshipout{\leftline{\vbox{\pagebody\makefootline}}}\advancepageno
}
\fi
%
%
%
\def\mayer{\vbox{\sl\centerline{Department of Physics 0319}%
\centerline{University of California, San Diego}
\centerline{9500 Gilman Drive}
\centerline{La Jolla, CA 92093-0319}}}
%
%

\def\pyiam{PHY-8958081}

%
%
\def\UCSD#1#2{\noindent#1\hfill #2%
\bigskip\supereject\global\hsize=\hsbody%
\footline={\hss\tenrm\folio\hss}}
%
%
\def\abstract#1{\centerline{\bf Abstract}\nobreak\medskip\nobreak\par #1}
%
%
%
%
\edef\tfontsize{ scaled\magstep3}
 \tfontsize  \tfontsize
 \tfontsize \font\titlei=cmmi10 \tfontsize
\font\titleis=cmmi7 \tfontsize \font\titleiss=cmmi5 \tfontsize
\font\titlesy=cmsy10 \tfontsize \font\titlesys=cmsy7 \tfontsize
\font\titlesyss=cmsy5 \tfontsize  \tfontsize
\skewchar\titlei='177 \skewchar\titleis='177 \skewchar\titleiss='177
\skewchar\titlesy='60 \skewchar\titlesys='60 \skewchar\titlesyss='60
%
%
%
%
%
\def\inv{^{\raise.15ex\hbox{${\scriptscriptstyle -}$}\kern-.05em 1}}
\def\lbar{{\lower.35ex\hbox{$\mathchar'26$}\mkern-10mu\lambda}}

%
%
%
%
\def\dsl{\,\raise.15ex\hbox{/}\mkern-13.5mu D} 
\def\delsl{\raise.15ex\hbox{/}\kern-.57em\partial}
\def\Ksl{\hbox{/\kern-.6000em\rm K}}
\def\Asl{\hbox{/\kern-.6500em \rm A}}
\def\Dsl{\hbox{/\kern-.6000em\rm D}} 
\def\Qsl{\hbox{/\kern-.6000em\rm Q}}
\def\gradsl{\hbox{/\kern-.6500em$\nabla$}}
%
%
\def\lspace{\ifx\answ\bigans{}\else\qquad\fi}
\def\lbspace{\ifx\answ\bigans{}\else\hskip-.2in\fi} 
%
%
\def\boxeqn#1{\vcenter{\vbox{\hrule\hbox{\vrule\kern3pt\vbox{\kern3pt
        \hbox{${\displaystyle #1}$}\kern3pt}\kern3pt\vrule}\hrule}}}
%
%
\def\mbox#1#2{\vcenter{\hrule \hbox{\vrule height#2in
\kern#1in \vrule} \hrule}}
%
%
%
%

   \def\CL{{\cal L}}
\def\CM{{\cal M}}  \def\CO{{\cal O}}

%
%
%
%
%
\def\del{\partial}

\def\bar#1{\overline{#1}}

\def\bra#1{\left\langle #1\right|}
\def\ket#1{\left| #1\right\rangle}

\def\darr#1{\raise1.5ex\hbox{$\leftrightarrow$}\mkern-16.5mu #1}

%
%
\def\half{{\textstyle{1\over2}}} 
\def\frac#1#2{{\textstyle{#1\over #2}}} 
%
%
%
%

\def\Tr{\mathop{\rm Tr}}

\def\GeV{{\rm GeV}}
\def\MeV{{\rm MeV}}

%
%
%
%

%
%
\def\ltap{\ \raise.3ex\hbox{$<$\kern-.75em\lower1ex\hbox{$\sim$}}\ }
\def\gtap{\ \raise.3ex\hbox{$>$\kern-.75em\lower1ex\hbox{$\sim$}}\ }
\def\gl{\ \raise.5ex\hbox{$>$}\kern-.8em\lower.5ex\hbox{$<$}\ }
\def\roughly#1{\raise.3ex\hbox{$#1$\kern-.75em\lower1ex\hbox{$\sim$}}}
%
%

%

%

\relax

\def \dtwo{{D_2^*}}
\def \done{{D_1}}
\def \dsone{D_{s1}}
\def \dstar{{D^*}}
\def \d{D}
\def \chdtwo{{D_2^{0*}}}
\def \chdone{{D_1^0}}
\def \pvec{{\done^\prime}}
\def \scal{{\d_0^*}}
\def \to{\rightarrow}
\def \({\left(}
\def \){\right)}
\def \[{\left[}
\def \]{\right]}
\def \chiral{SU(3)_L\times SU(3)_R}
\def \genhad{H_a^{\mu_1\dots\mu_k}}
\def \vslash{v\hskip-0.5em /}

\def \sproj{{(1+\vslash)\over 2\sqrt{2}}}
\def \gamf{\gamma_5}
\def \gmu{\gamma_\mu}
\def \fpi{f_\pi}
\def \lchi{\Lambda_\chi}
\def \dms{\delta m_S}
\def \dmt{\delta m_T}
\def \brownmuck{light degrees of freedom\ }
\def \clebsch#1{\sqrt{\frac#1}}
\def \ppi{\vec{p}_\pi}
\def \im{{\rm i}}

\def\mayer{\vbox{\sl\centerline{Department of Physics 0319}%
\centerline{University of California, San Diego}
\centerline{9500 Gilman Drive, La Jolla, CA 92093-0319}}}

\centerline{{\titlefont{Strong Decays of Excited Heavy Mesons}}}
\bigskip
\centerline{{\titlefont{In Chiral Perturbation Theory}}
\footnote{\raise1ex
\hbox{\hskip-.3em *}}{Work supported by Department of Energy contracts
DE-AC03-76SF00515 and DE-FG03-90ER40546, and by National Science
Foundation grant \pyiam.}}
\bigskip\bigskip
\centerline{Adam F. Falk}
\medskip
\vbox{\sl\centerline{Stanford Linear Accelerator Center}
\centerline{Stanford University, Stanford, California 94309}}
\medskip
\centerline{and}
\medskip
\centerline{Michael Luke}
\medskip
\mayer
\bigskip
\vfill
\abstract{We construct an effective Lagrangian describing the interaction
of soft pions and kaons with mesons containing a heavy quark and light
degrees of freedom in an orbital $p$ wave.  The formalism is easily
extended to heavy mesons and baryons in arbitrary excited states.
We calculate the leading contributions to the
strong decays $\dtwo\to\d\pi$, $\dtwo\to\dstar\pi$
and $\done\to\dstar\pi$.  We confirm the relations between the rates
previously obtained by Isgur and Wise using heavy quark symmetry,
and find that the absolute widths are
consistent with na\"\i ve power counting.
We also estimate the
branching ratios for the two pion decays $\dtwo\to\dstar\pi\pi$,
$\done\to\dstar\pi\pi$ and $\done\to\d\pi\pi$, which
are dominated by pole graphs.  Our predictions depend on the masses
and widths of the as yet unseen scalar-pseudovector $p$-wave doublet.
Heavy quark spin symmetry predicts
$\Gamma(\dtwo\to\dstar\pi\pi):
\Gamma(\done\to\dstar\pi\pi):\Gamma(\done\to\d\pi\pi)=3:1:2$, but this
relation is badly violated in practice because $1/M$ effects arising
purely from kinematics are large.
}
\bigskip
\UCSD{UCSD/PTH 92-14, SLAC-PUB-5812}{May 1992}

\newsec{Introduction}

The interactions of the octet of pseudogoldstone bosons with
hadrons containing a single heavy quark are constrained by two
independent symmetries: spontaneously broken chiral $\chiral$
and heavy quark spin-flavour $SU(2N_h)$ \ref\hqss{N.~Isgur
and M.~B.~Wise,
Phys.~Lett.~B232 (1989) 113, Phys.~Lett.~B237 (1990) 527}.
One may implement both of
these symmetries by constructing a ``heavy-light'' chiral lagrangian,
in which one performs a simultaneous expansion in the momenta of the
pseudogoldstone bosons and the inverse masses of the heavy hadrons.
Such a lagrangian has been described in refs.~\ref\wise{M.~B.~Wise,
Caltech preprint CALT-68-1765 (1992)}--%
\nref\donoghue{G.~Burdman and J.~Donoghue, U.~Mass.~preprint UMHEP-365
(1992)}%
\nref\dudes{T.-M.~Yan, H.-Y.~Cheng, C.-Y.~Cheung, G.-L.~Lin,
Y.~C.~Lin and H.-L.~Yu, Cornell preprint CLNS 92/1138 (1992)}%
\ref\peter{P.~Cho, Harvard preprint HUTP-92/A014 (1992)}
for heavy
hadrons with the light degrees of freedom in the ground state.
We begin by briefly reviewing this construction.

The lagrangian is written in terms of the usual exponentiated matrix of
pseudogoldstone bosons,
\eqn\pseudo{\xi=\exp\(\im\CM/\fpi\),\qquad\Sigma\equiv\xi^2}
where
\eqn\cm{\CM=\pmatrix{{\textstyle{1\over\sqrt{2}}}\pi^0+{\textstyle
    {1\over\sqrt{6}}}\eta&\pi^+&K^+\cr\pi^-
    &{\textstyle{-{1\over\sqrt{2}}}}\pi_0+
    {\textstyle{1\over\sqrt{6}}}\eta&K^0\cr
    K^-&\bar K^0&-{\textstyle\sqrt
    {2\over 3}}\eta}}
and $\fpi\approx135\,\MeV$.  Under chiral $\chiral$, the field $\xi$
transforms as $\xi\to L\xi U^\dagger = U\xi R^\dagger$, where $U$ is a
matrix which depends on the fields $\CM$, while $\Sigma$ transforms
more simply as $\Sigma\to L\Sigma R^\dagger$.  The ground state heavy
mesons consist of a doublet under heavy quark spin symmetry,
containing the pseudoscalar meson $P$ and the vector meson $P^*$; these
also transform under the unbroken flavour $SU(3)$ as an antitriplet.  (We
take our heavy mesons always to contain a heavy quark rather than an
antiquark.)   We represent these fields in the usual way by a $4\times4$
Dirac matrix,
\eqn\defineh{H_a=\sproj\[P^{*\mu}_a\gmu - P_a\gamf\]\,.}
We have absorbed factors of $\sqrt{2M_P}$ and $\sqrt{2M_P^*}$ into
the definition of the heavy fields, so they have mass dimension $3/2$
(our normalisation differs slightly from that of ref.~\wise; our fields
are normalised to 1, not to 2).  To recover the correct relativistic
normalisation, we multiply amplitudes by $\sqrt{2M}$ for each external
heavy meson.

The pseudogoldstone bosons couple to the heavy fields through the
covariant derivative
\eqn\covder{D_{ab}^\mu\equiv\delta_{ab}\del^\mu+V_{ab}^\mu
    =\delta_{ab}\del^\mu+\half\(\xi^\dagger
    \del^\mu\xi+\xi\del^\mu\xi^\dagger\)_{ab}}
and the axial vector field
\eqn\axial{A_{ab}^\mu={\textstyle {\im\over 2}}\(\xi^\dagger\del^\mu\xi
    -\xi\del^\mu\xi^\dagger\)_{ab}\,.}
Under $\chiral$,
\eqn\htrans{H_a\to U_{ab}H_b\,,\qquad (D_\mu H)_a\to U_{ab}(D_\mu H)_b\,,
    \qquad A^\mu_{ab}\to U_{ac}A^\mu_{cd}U^{\dagger}_{db}\,.}
At leading order in the momentum expansion, the lagrangian is written
in terms of these fields as
\eqn\original{\eqalign{\CL&={\fpi^2\over 8}\del^\mu\Sigma_{ab}
    \del_\mu\Sigma^\dagger_{ba}-\Tr\[\bar H_a\im v\cdot D_{ba} H_b\]
    +g\Tr\[\bar H_a H_b\,\Asl_{ba}\gamma_5\]\cr &\qquad
    +\lambda_0\[m_{\rm q}\Sigma+m_{\rm q}
    \Sigma^\dagger\]_{aa}+\cdots\,,\cr}}
where the traces are over Dirac indices and we keep the $SU(3)$ flavour
indices $a$, $b$ explicit.
The ellipses denote terms higher order in the derivative
expansion, terms suppressed by powers of $1/M$, and additional
explicit $\chiral$ violating terms proportional to the quark mass matrix
\eqn\mq{m_{\rm q}=\pmatrix{m_{\rm u}&0&0\cr 0&m_{\rm d}&0\cr
    0&0&m_{\rm s}}\,.}

\newsec{Excited States and Reparameterisation Invariance}

We would now like to consider the form of such a lagrangian for heavy
mesons in an excited state.
In the limit that
the heavy quark mass $M$ is taken to infinity the light degrees of
freedom carry a well-defined angular momentum, flavour and spectrum of
excitations.
In general, the \brownmuck in a heavy
meson are in a state with half-integral
angular momentum $j$ and parity $P$,
corresponding
to two degenerate heavy mesons of spin $j\pm\half$ and parity $-P$ (since
quarks and antiquarks have opposite parity).  We may describe both
states by a more complicated analogue of the $H_a$ matrix \defineh,
the traceless, symmetric Lorentz tensor
\eqn\general{\genhad\,,\quad k=j-1/2\,,}
satisfying the conditions
\eqn\genconds{v_{\mu_1}\genhad=\gamma_{\mu_1}\genhad=0\,.}
Under Lorentz transformations,
\eqn\lorentz{\genhad\to D(\Lambda)
    \Lambda^{\mu_1}_{\nu_1}\dots\Lambda^{\mu_k}_{\nu_k}
    H_a^{\nu_1\dots\nu_k}D^\dagger(\Lambda)\,,}
where $D(\Lambda)$ is an element of the $4\times 4$ matrix
representation of the Lorentz group, while under spatial rotations
$\widetilde\Lambda$ of the heavy quark,
\eqn\hqrot{\genhad\to D(\widetilde\Lambda)\genhad\,.}
The general form for $\genhad$ has been derived in
ref.~\ref\falk{A.~F.~Falk,
SLAC preprint SLAC-PUB-5689 (1991), to appear in Nucl. Phys.~B};
for \brownmuck with parity $(-1)^{j-1/2}$ we have the doublet of
states\footnote{$^\star$}{We use here the
particle data book convention of labeling states
with a subscript for their spin, and adding a superscript ``$^*$''
if the spin-parity is in the series $J^P=0^+, 1^-, 2^+,\dots$ .}
$Q^*_{j+1/2}$ and $Q_{j-1/2}$,
\eqn\bmone{\eqalign{\genhad=&\sproj
    \left\{(Q_{j+1/2}^*)_a^{\mu_1\dots\mu_{k+1}}
    \gamma_{\mu_{k+1}} - {\textstyle\sqrt{{2k+1\over k+1}}}\gamf
    (Q_{j-1/2})_a^{\nu_1\dots\nu_k}\right.\cr
    &\quad\[g^{\mu_1}_{\nu_1}\dots g^{\mu_k}_{\nu_k}
    -{\textstyle{1\over 2k+1}}\gamma_{\nu_1}\(\gamma^{\mu_1}-v^{\mu_1}\)
    g^{\mu_2}_{\nu_2}\dots g^{\mu_k}_{\nu_k} - \cdots \right.\cr
    &\qquad\left.\left.-{\textstyle{1\over 2k+1}}
    g^{\mu_1}_{\nu_1}\dots g^{\mu_{k-1}}_{\nu_{k-1}}\gamma_{\nu_k}
    \(\gamma^{\mu_k}-v^{\mu_k}\)\]\right\}\,,}}
while for parity $(-1)^{j+1/2}$ we have $Q_{j+1/2}$ and $Q^*_{j-1/2}$,
\eqn\bmtwo{\eqalign{\genhad=&\sproj
    \left\{(Q_{j+1/2})_a^{\mu_1\dots\mu_{k+1}}
    \gamf\gamma_{\mu_{k+1}} - {\textstyle \sqrt{{2k+1\over k+1}}}
    (Q_{j-1/2}^*)_a^{\nu_1\dots\nu_k}\right.\cr
    &\quad\[g^{\mu_1}_{\nu_1}\dots g^{\mu_k}_{\nu_k}
    -{\textstyle{1\over 2k+1}}\gamma_{\nu_1}\(\gamma^{\mu_1}+v^{\mu_1}\)
    g^{\mu_2}_{\nu_2}\dots g^{\mu_k}_{\nu_k} - \cdots\right.\cr
    &\qquad\left.\left.-{\textstyle{1\over 2k+1}}
    g^{\mu_1}_{\nu_1}\dots g^{\mu_{k-1}}_{\nu_{k-1}}\gamma_{\nu_k}
    \(\gamma^{\mu_k}+v^{\mu_k}\)\]\right\}\,.}}
For simplicity, we will restrict ourselves in the rest of this paper
to the lowest lying $p$-wave excitations; we have included the complete
expressions \bmone\ and \bmtwo\ to make it clear that the extension of
this formalism to arbitrary excited heavy mesons is cumbersome but
straightforward.  (Using the formalism of ref.~\falk, one could include
excited heavy baryons as well.)
In the quark model, these $p$-wave states
correspond to \brownmuck with orbital angular
momentum $\ell=1$, and hence with total spin $j=\half$ or $j=\frac32$.
For the $\d$ system, these are the (as-yet unobserved)
$J^P=0^+, 1^+$ doublet $\scal$ and
$\pvec$,
\eqn\dbone{S_a=\sproj\(\pvec^\mu\gmu\gamf-\scal\)\,,}
and the $J^P=1^+, 2^+$ doublet $\done$ and $\dtwo$,
\eqn\dbtwo{T^\mu_a=\sproj\left\{\dtwo^{\mu\nu}\gamma_\nu
    -\sqrt{\frac 32}\done^\nu\gamf\[g^\mu_\nu-\frac 13 \gamma_\nu
    \(\gamma^\mu-v^\mu\)\]\right\}}
(we add a prime to distinguish the two pseudovector states).
We identify the neutral members of this multiplet as the
$\done(2420)^0$ and the $\dtwo(2460)^0$ \ref\pdb{Particle Data Group,
Phys.~Lett.~B232 (1990) 1}.
Including these states along with the ground state mesons, the kinetic
piece of the chiral Lagrangian is given by
\eqn\lkin{\eqalign{\CL_{\rm kin}=&-\Tr\[\bar H_a \im v\cdot D_{ba} H_b\]
    +\Tr\[\bar S_a\(\im v\cdot D_{ba}-\dms\delta_{ba}\) S_b\]\cr
    &\qquad+\Tr\[\bar T^\mu_a \(\im v\cdot D_{ba}-\dmt\delta_{ba}\)
    T_{\mu b}\]\,,\cr}}
where the residual masses $\dms=M_\scal-M_D=M_\pvec-M_D$ and
$\dmt=M_\done-M_D=M_\dtwo-M_D$
are defined in the heavy quark limit, where the doublets are
degenerate \ref\epic{A.~F.~Falk,
M.~Luke and M.~Neubert, SLAC and UCSD preprint SLAC-PUB-5771
and UCSD/PTH 92-09 (1992)}.

In general, one must include all terms in a chiral Lagrangian which are
not forbidden by symmetries of the effective theory.
Hence one might be tempted to write down a mixing term of the form
\eqn\evilterm{\Tr\[\bar H_a\(\im D_\mu T^\mu\)_a\]}
(recall that in the effective theory, $D_\mu T^\mu\ne 0$; the
transversality condition is $v_\mu T^\mu=0$).
However, such a term is forbidden because it is not invariant under
redefinitions of the velocity $v^\mu$ \ref\lm{M.~Luke and A.~V.~Manohar,
UCSD preprint UCSD/TH 92-15 (1992)}.
Recall that the definition of the velocity $v^\mu$ of a heavy field of
mass $M$ is somewhat arbitrary, in that  we could
equally well choose a slightly different velocity $v'^\mu=v^\mu-q^\mu/M$,
where $q\cdot v=q^2/2M$ to ensure $v'^2=1$,
and shift the residual momentum by $q^\mu$:
\eqn\momentum{P^\mu=Mv^\mu+k^\mu=Mv'^\mu+k^\mu+q^\mu.}
For a heavy scalar $\phi$ or vector field $A^\mu$, this corresponds to
the transformation
\eqn\reparam{\eqalign{v^\mu&\to v^\mu-\frac{1}{M} q^\mu\,,\cr
    \phi&\to e^{iq\cdot x}\phi\,,\cr
    A^\mu&\to\[g^{\mu\nu}+\frac{1}{M} v^\mu
    q^\nu+\CO\(\frac{1}{M^2}\)\] e^{iq\cdot x}A_\nu\,.}}
Under the shift \reparam,
\eqn\notinv{\Tr\[\bar H_a\(\im D_\mu T^\mu\)_a\]\to
    \Tr\[\bar H_a\( (\im D_\mu-q_\mu) T^\mu\)_a\]+\CO(1/M)\,,}
so the term \evilterm\ is forbidden.

The single pion transitions between states in the same heavy spin
doublet are given by terms in the effective lagrangian analogous to
the $g$ coupling in eq.~\original:
\eqn\onepiona{\CL_{1\pi}=g\Tr\[\bar H_a H_b\,\Asl_{ba}\gamf\]
    +g'\Tr\[\bar S_a S_b\,\Asl_{ba}\gamf\]
    +g''\Tr\[\bar T^\mu_a T_{\mu b}\,\Asl_{ba}\gamf\]\,,}
while the single pion transitions between doublets, again to lowest
order in the derivative expansion, are given by
\eqn\onepionb{\CL_{s}=f'\Tr\[\bar S_a T^\mu_b A_{\mu ba}\gamf\] +
    f''\Tr\[ \bar H_a S_b\,\Asl_{ba}\gamf\]+{\rm h.c.}\,.}
These correspond to $s$-wave transitions; however the analogous $s$-wave
transitions $T^\mu\to H\pi$ are forbidden by heavy quark spin symmetry
\ref\marknathan{N.~Isgur and M.~B.~Wise, Phys.~Rev.~Lett.~D66 (1991)
1130}, and indeed the term $\Tr\[\bar H_a T^\mu_b A_{\mu ba}\gamf\]$
vanishes.  These decays must then proceed through $d$-waves, which
are suppressed by one derivative in the chiral lagrangian:
\eqn\pidecay{\CL_{d}={h_1\over\lchi}\Tr\[\bar H_a T^\mu_b\(\im D_\mu\,
    \Asl\)_{ba}\gamf\]+{h_2\over\lchi}\Tr\[\bar H_a T^\mu_b\(\im\dsl
    A_\mu\)_{ba}\gamf\] + {\rm h.c.}\,,}
where $\lchi$ is some momentum scale characterising the convergence
of the derivative expansion.  From previous experience with chiral
Lagrangians, we expect $\lchi\simeq 1\ \GeV$ \ref\bighoward{See, for
example, H.~Georgi, {\sl Weak Interactions and Modern Particle Theory},
Benjamin/Cummings Publishing Co., Menlo Park, CA (1984)}, and so
we expect the $T^\mu$ states to be much narrower than the $S$ states,
simply from power counting.
Note that the symmetry \reparam\ also forbids couplings such as
$\Tr\[(\im D_\mu \bar H)_a T^\mu_b\, \Asl_{ba}\gamf\]$, with derivatives
acting on the heavy fields, at this order in $1/M$.

Following the authors of ref.~\dudes, who obtained an estimate of $g$,
we may estimate the couplings $g', g''$ and $f'$ in the
nonrelativistic quark model by evaluating matrix elements of the axial
current between the appropriate states.  This requires the assumption
that the pseudogoldstone bosons couple only to the spin of the brown
muck, and not to the orbital angular momentum.
We note that the nonrelativistic quark model may not provide a
very appropriate
description of these excited states, as the mass splitting from the
ground state is of the order of several hundred MeV, comparable to the
mass of the constituent light quark.  Hence we should probably regard our
estimates of the couplings primarily as an indication of what are likely
to be reasonable values for these parameters.

In the nonrelativistic quark model,
the $S_a$ and $T^\mu_a$ mesons have the
\brownmuck in the same excited radial wavefunction, and we may decompose
physical states into the eigenstates $\ket{s_H, m_\ell, s_\ell}$
of the $z$ components of heavy quark spin $s_H$,
angular momentum of the \brownmuck $m_\ell$ and
light quark spin $s_\ell$.
In particular, we decompose the $m=0$ states of the
$\dtwo$ and the $\pvec$ as
\eqn\qmdtwo{\eqalign{\ket{\dtwo(m=0)}=&\clebsch{13}\ket{\half,0,-\half}
+\clebsch{16}\ket{\half,-1,\half}\cr&\quad+\clebsch{16}
\ket{-\half,1,-\half}
+\clebsch{13}\ket{-\half,0,\half}}}
and
\eqn\qmpvec{\eqalign{\ket{\pvec(m=0)}=&\clebsch{16}\ket{\half,0,-\half}
-\clebsch{13}\ket{\half,-1,\half}\cr&\quad+\clebsch{13}
\ket{-\half,1,-\half}
-\clebsch{16}\ket{-\half,0,\half}.}}
Consider the matrix element of the axial current $(j_5^i)^\mu$ between
these states.  In the nonrelativistic quark model,
\eqn\axcurr{(j^{1+\im2}_5)^3=-g_A u^\dagger \sigma^3 d,}
where we take $g_A=0.75$ as suggested by the chiral quark model
\ref\mg{A.~Manohar and H.~Georgi, Nucl.~Phys.~B234 (1984) 189}
(this reproduces
the correct value of $g_A$ in the nucleon).  Thus we obtain
\eqn\nrqm{
    \bra{\vphantom{\pvec}\dtwo(m=0)}\int d^3x\,
    (j^{1+\im2}_5)^3\ket{\pvec(m=0)}={2\sqrt{2}\over3}g_A\,.}
In the chiral lagrangian, the $f'$ coupling in \onepionb\ gives a
contribution to the axial current of
\eqn\jmufive{(j^i_5)^\mu=-f'\Tr\[\bar S_a T^\mu_b\gamf
T^i_{ba}\]+\dots\,.}
In the limit of zero momentum transfer, this term dominates the matrix
element \nrqm\ and we find
\eqn\chiralmat{\bra{\vphantom{\pvec}\dtwo(m=0)}
    \int d^3x\,(j^{1+\im2}_5)^3\ket{\pvec(m=0)}=-f'\epsilon^*_\mu
    \eta^{\mu3}=-\sqrt{2\over3}f'\,,}
where $\epsilon^\mu$ and $\eta^{\mu\nu}$ are respectively the $m=0$
polarisation states of the $\pvec$ and
$\dtwo$.  Equating the expressions \nrqm\ and \chiralmat, we find
\eqn\fprime{|f'|={2\over\sqrt3}g_A=0.87\,.}
The phase of $f'$ is not determined by this procedure;
however this will not matter as only the modulus $|f'|^2$ will appear
in the widths which we will compute.
Similarly, we may obtain estimates of the transition rates within
multiplets,
\eqn\useless{g=g_A\,,\qquad g'=\frac13 g_A\,,\qquad g''=g_A\,,}
where the phases may in this case be fixed by the heavy quark symmetry
relation ${\bf S}_h^z\ket{D^*(m=0)}=\textstyle{1\over2}\ket{D}$, and
analogously for the excited doublets.
However, the coupling constants $g'$ and $g''$ are not particularly
useful, as the
corresponding single pion decays are most
probably kinematically forbidden
\ref\godfrey{S.~Godfrey and N.~Isgur, Phys.~Rev.~D32 (1985) 189\semi
S.~Godfrey and R.~Kokoski, Phys.~Rev.~D43 (1991) 1679}.

\newsec{Single Pion Decays}

There are four possible single pion transitions between two heavy spin
doublets; relations between the amplitudes follow from the heavy quark
spin symmetry.  These
have already been worked out explicitly for $\dtwo$ and $\done$
decays \marknathan\ref\rosner{J.~Rosner, Comm.~Nucl.~Part.~Phys.~16
(1986) 109}, and those results follow immediately from our
formalism.  In addition, with the chiral
lagrangian we may easily correct for one class of $1/M$ corrections
in the widths
by using the true particle masses in the phase space integrals.  Since
the rate for $d$-wave decays is proportional to the fifth power of
the pion momentum, this is likely
to be the leading $1/M$ correction.  Explicitly, we find
\eqn\dwave{\eqalign{
  &\Gamma(\chdtwo\to\d^+\pi^-)={1\over 15\pi} \(M_\d\over M_\dtwo\)
  {h^2\over\lchi^2} {|\ppi|^5\over f_\pi^2} = 5.51\times 10^7
  {h^2\over\lchi^2}\,,\cr
  &\Gamma(\chdtwo\to\dstar^+\pi^-)={1\over 10\pi}
  \(M_\dstar\over M_\dtwo\)
  {h^2\over\lchi^2} {|\ppi|^5\over f_\pi^2} = 2.03\times 10^7
  {h^2\over\lchi^2}\,,\cr
  &\Gamma(\chdone\to\dstar^+\pi^-)={1\over 6\pi}
  \(M_\dstar \over M_\done\)
  {h^2\over\lchi^2} {|\ppi|^5\over f_\pi^2} = 2.05\times 10^7
  {h^2\over\lchi^2}\,,}}
where $\ppi$ is the momentum of the pion emitted in the decay, and
$h\equiv |h_1+h_2|$.  The full one pion width are $3/2$ times these
because of the $D^0\pi^0$ channel.  From eq.~\dwave\ we reproduce
the result of Isgur and Wise,
\eqn\ratio{{\Gamma(\chdtwo\to\d^+\pi^-)\over\Gamma
(\chdtwo\to\dstar^+\pi^-)}
    = 2.7\,,}
which compares very well with the experimental ratio $2.4\pm 0.7$
\pdb.
We may use these results to gain some confidence in the
validity of our derivative expansion.  Assuming the total $\dtwo$ width
of $19\pm 7\,\MeV$
to be saturated by the one pion mode (as we will
show in the next section, the two pion width is sufficiently small that
this is a reasonable assumption), we find
\eqn\worksdontit{{h^2\over\lchi^2}\approx{1\over (2\ {\rm GeV})^2}\,,}
which is consistent with our na\"\i ve estimate.
  This also gives
us a prediction for the $\chdone$ single pion width,
\eqn\donewidth{\Gamma(\chdone\rightarrow D^{*+}\pi^-+D^{*0}\pi^0)\approx
    7\,\MeV,}
which is significantly smaller than the measured total width of
$20^{+9}_{-5}\,\MeV$.  As has been suggested \marknathan, this is
undoubtedly
due to mixing (at order $1/M$) of the $\done$ with the substantially
broader $\pvec$.

The $\scal$ and $\pvec$ decay through $s$-wave pion emission and
consequently are very broad;  from eq.~\onepionb\ we obtain
\eqn\swave{\eqalign{\Gamma(\scal\to\d\pi^-)=&
    {|f''|^2\over 2\pi f_\pi^2}\({M_\d\over M_\scal}\)
    (M_\scal-M_\d)^2\left[(M_\scal-M_\d)^2-m_\pi^2\right]^{1/2}\cr
    \Gamma(\pvec\to\dstar\pi^-)=&
    {|f''|^2\over 2\pi f_\pi^2}\({M_\dstar\over M_\pvec}\)
    (M_\pvec-M_\dstar)^2\cr
&\qquad\times\left[(M_\pvec-M_\dstar)^2-m_\pi^2\right]^{1/2}.}}
Since these states have not been observed, we must use quark model
estimates for their masses.  Taking $M_\scal=M_\pvec=2.4\ \GeV$ \godfrey,
we have
\eqn\swavepred{\eqalign{&\Gamma(\scal\to\d\pi^-)=|f''|^2[980\,\MeV]\cr
&\Gamma(\pvec\to\dstar\pi^-)=|f''|^2[400\,\MeV].}}
Again, the full one pion widths are 3/2 times these because of
the $\pi^0$ channel.  These widths are very sensitive to the value used
for the mass of the states; for $M_\scal=M_\pvec=2.3\,\GeV$ we find
charged pion widths of $|f''|^2[540\,\MeV]$ and $|f''|^2[160\,\MeV]$,
respectively.

\newsec{Two Pion Decays}

Like the single pion $T^\mu\rightarrow H\pi$ decays, the contact terms
mediating $T^\mu\rightarrow H\pi\pi$, such as $\Tr\[\bar H_a T^\mu_b
A_{\mu bc}\,\Asl_{ca}\]$,
are dimension five and are
suppressed by one power of $\lchi$ in the derivative expansion.
We therefore expect that these decays will be dominated
by pole graphs in which there is an intermediate $\pvec$ or $\scal$
which is close to its mass shell.
This raises the interesting possibility that the two pion widths could
be comparable to the single pion widths (as is observed,
for example, in the decay $K_2^*(1430)\rightarrow K^*(892)+$pions).
The two pion width is given by
\eqn\twopiwidth{\Gamma_{2\pi}=\int {1\over (2\pi)^3} {1\over 8M'}
    |{\cal A}(E_1,E_2)|^2 dE_1 dE_2\,,}
where the amplitude ${\cal A}$ is a function of the energies $E_1$ and
$E_2$ of the outgoing pions, and the masses $M'=(M_\dtwo, M_\done)$ and
$M=(M_\d,M_\dstar)$ are those respectively of
the initial and final heavy mesons.
Kinematics restricts $E_2$ to the region
\eqn\limits{\bar E_2(E_1)-{g(E_1)\over M}<E_2<\bar E_2(E_1)
    +{g(E_1)\over M}\,,}
where
\eqn\defineg{\eqalign{&\bar E_2(E_1)\equiv M'-M-E_1\,,\cr
&g(E_1)\equiv \sqrt{(E_1^2-m_\pi^2)[(M'-M-E_1)^2-m_\pi^2]}\,.}}
Hence the amplitude can be expressed approximately as a function only
of $E_1$,
\eqn\almost{{\cal A}(E_1,E_2)\simeq {\cal A}\(E_1,\bar E_2(E_1)\)}
up to corrections of order $1/M$.
The integral over $E_2$ then just brings in a factor of the width
of the integral, $2g(E_1)/M$.
Because of the poles in the intermediate $\pvec$ and $\scal$
propagators, their widths must be included in our expressions.
The imaginary part of the propagator of this resonance is
\eqn\improp{\Gamma_{\rm int}(p\cdot v)={|f''|^2\over 2\pi f_\pi^2}
    {M\over M_{\rm res}}
    (M_{\rm res}-M+p\cdot v)^2\left[(M_{\rm res}-M+p\cdot v)^2-m_\pi^2
    \right]^{1/2}\,,}
where $p$ is the residual momentum flowing through the line
and $M_{\rm res}=M_\scal$ or $M_\pvec$.
For $p\cdot v\simeq 0$, this reduces to the usual Breit-Wigner formula.
However, because these states are so broad we must include the full
momentum dependence of the width in the denominator.  It is convenient
to extract from the $|{\cal A}(E_1,\bar E_2)|^2$ the function
\eqn\yetanother{\openup1\jot \eqalign{F(E_1)=&{E_1^2[(M_\done-M_\d-E_1)^2
    -m_\pi^2]\over (E_1-[M_\scal-M_\d])^2
    +\Gamma_{\rm int}(E_1-[M_\scal-M_\d])^2/4}\cr
    &\qquad + {(M_\done-M_\d-E_1)^2[E_1^2-m_\pi^2]\over
    [(M_\done-M_\scal-E_1)]^2+\Gamma_{\rm
    int}(M_\done-M_\scal-E_1)^2/4}\,,}}
where there are two terms because the pions may be emitted in either
order (the cross terms in $|{\cal A}|^2$ integrate to zero).
Then the partial width is given by
\eqn\thewidth{\Gamma_{\pi^-\pi^0}=
    {\alpha\over 4(2\pi)^3}{|f' f''|^2\over f_\pi^4}
    \int F(E_1)g(E_1) dE_1\,,}
where $\alpha=2/9$ for $\chdone\to\dstar\pi^-\pi^0$, $\alpha=4/9$ for
$\chdone\to\d\pi^-\pi^0$ and $\alpha=2/3$ for
$\chdtwo\to\dstar\pi^-\pi^0.$
There are also decays to a neutral charmed hadron and a $\pi^+\pi^-$
pair which occur with the same amplitude
(since the final pions are in an antisymmetric wave function, the $I=0$
$\pi^0\pi^0$ mode is forbidden).  Hence the full two pion
widths are twice those given in eq.~\thewidth.
Our predictions for the two pion widths depend on several unknown
parameters:  the masses and widths of the as yet unobserved $\scal$
and $\pvec$, as well as on the couplings $f'$ and $f''$.  In \fig\graphs{
Full two pion widths for $\dtwo$ and $\done$ as functions of
the $\pvec$ width, for $M_\scal=M_\pvec=2300\,\MeV$ and $2400\,\MeV$.
Note that for $M_\pvec=2300\,\MeV$ the $\dtwo$ partial width
is nonzero as the $\pvec$ width goes to zero, since the $\pvec\pi$
intermediate state
may be produced on shell.  In this limit the $\dtwo$ two pion
width approaches
the $\dtwo\rightarrow\pvec\pi$ partial width.}
we plot the total two-pion decay widths for $\done\to\dstar\pi
\pi$, $\done\to\d\pi\pi$ and $\dtwo\to\dstar\pi\pi$ as
functions of $|f''|$, or equivalently as functions of the $\pvec$
width, assuming
the nonrelativistic quark model prediction \fprime\ for $f'$.
Variations in $f'$ just change the overall normalisations, but not the
shapes, of the plots.  Note that in the heavy quark limit, the widths
would satisfy
\eqn\inthelimit{\Gamma(\dtwo\to\dstar\pi\pi)_{M\rightarrow\infty}:
\Gamma(\done\to\dstar\pi\pi)_{M\rightarrow\infty}:
\Gamma(\done\to\d\pi\pi)_{M\rightarrow\infty}=3:1:2\,,}
but because of the sensitive dependence of eq.~\thewidth\ on the
masses, this relation is badly violated.  So although in
this limit our approach simply reproduces the general
results of ref.~\marknathan, it has the advantage of being able to take
into account the large but calculable $1/M$ symmetry
breaking effects which arise purely from kinematics.  Since the
precise form of the $1/M$ effects depends on the fact that the
decay is dominated by pole graphs, its exact form could not be
guessed (unlike the $|\ppi|^5$ behaviour for the single pion decays).

One might also think to apply this analysis to the strong transitions
of excited strange, charmed mesons.  Indeed, at least one such state,
the $\dsone$, as already been observed \pdb.  However, such decays are
severely constrained by the combination of phase space and the heavy
quark limit.  If the outgoing $D$ meson is not strange, there must be a
$K$ meson in the final state,  but $\dsone\to\d K$ is prohibited in the
heavy quark limit, while $\done\to\dstar K$ is barely possible
kinematically and hence severely suppressed.  As for decays to ground
state $D_s$ mesons, there is not enough energy to emit the isospin-0
$\eta$, while the decay to two pions in an isospin-0 state is induced
by our effective lagrangian only at the one loop level.  The strong
decays of the $\dsone$ are thus most likely mediated by operators which
are subleading in the mass expansion.  Although one might expect the
current mass of the strange quark to induce larger $1/M$ corrections in
the $D_s$ system than in the $D^+$ and $D^0$, this suppression might
help explain the relatively narrow width ($<5\,\MeV$) observed for the
$\dsone$.

This formalism could also be applied to semileptonic decays
from a $B$ meson to an excited $D$ plus soft pions, as has been
done for decays to ground state $D$ mesons \ref\llw{C.~L.~Y.~Lee,
M.~Lu and M.~B.~Wise, Caltech preprint
CALT-68-1771 (1992)}.  There may also be significant
contributions to the decay $B\rightarrow D\pi\ell\bar\nu$, in
which the $B$ first decays semileptonically to
an excited $D$, which then decays strongly to a $D$ or $D^*$ and a
pion.  We are currently studying these processes.

Finally, we point out that the same heavy-light chiral
lagrangian could be used as well to describe the strong transitions of
excited bottom mesons.  In fact, we would expect $1/M$ corrections to
be considerably smaller than in the case of charm.  However these
states have not yet been produced and studied, and their masses, to
which the decay rates are so sensitive, are not known.

\vfil\break\eject
\centerline{\bf{Acknowledgements}}
\bigskip
It is a pleasure to thank Mark Wise both for many helpful discussions and
for the hospitality of the Caltech physics department, where portions
of this work were completed.  We are grateful to Ben Grinstein for
suggesting that an analysis such as this one might prove interesting.
We also thank Martin Savage and Aneesh Manohar for useful discussions.

\listrefs
\listfigs
\bye